\documentclass[12pt,epsf]{iopart}

\usepackage[dvips]{graphicx}
\usepackage{iopams}
\usepackage{epsfig}

\newcommand{\tfrac}[2]{\mbox{$\frac{#1}{#2}$}}
\font\capfont=cmbx12 at 50 pt % or yinit, or...?
\newbox\capbox \newcount\capl \def\a{A}
\def\docappar{\medbreak\noindent\setbox\capbox\hbox{%
\capfont\a\hskip0.15em}\hangindent=\wd\capbox%
\capl=\ht\capbox\divide\capl by\baselineskip\advance\capl by1%
\hangafter=-\capl%
\hbox{\vbox to8pt{\hbox to0pt{\hss\box\capbox}\vss}}}
\def\cappar{\afterassignment\docappar\noexpand\let\a }
\begin{document}

\newcommand{\ee}{{\rm e}}
\newcommand{\dd}{{\rm d}}
\newcommand{\p}{\partial}
\newcommand{\ok}{n^{\rm w}}   %{n_w}
\newcommand{\pok}{n^{\rm i}}  %{n_i}
\newcommand{\dok}{n}
\newcommand{\calT}{{\cal T}}
\newcommand{\bJ}{\bar{J}}
\newcommand{\bj}{\bar{j}}
\newcommand{\Jin}{J^{\rm{in}}}
\newcommand{\Jout}{J^{\rm{out}}}
\newcommand{\bJin}{\bar{J}^{\rm{in}}}
\newcommand{\bJout}{\bar{J}^{\rm{out}}}
\newcommand{\rhoin}{\rho^{\rm{in}}}
\newcommand{\rhoout}{\rho^{\rm{out}}}
\newcommand{\rhooutc}{\rho^{\rm{out}}_{\rm c}}
\newcommand{\ac}{a_{\rm c}}

\newcommand{\la}{\langle}
\newcommand{\ra}{\rangle}
\newcommand{\beq}{\begin{equation}}
\newcommand{\eeq}{\end{equation}}
\newcommand{\bea}{\begin{eqnarray}}
\newcommand{\eea}{\end{eqnarray}}
\def\lsim{\:\raisebox{-0.5ex}{$\stackrel{\textstyle<}{\sim}$}\:}
\def\gsim{\:\raisebox{-0.5ex}{$\stackrel{\textstyle>}{\sim}$}\:}

%\numberwithin{equation}{section}

\thispagestyle{empty}
\title[Frozen shuffle update for TASEP]{
Frozen shuffle update\\
for an asymmetric exclusion process\\
on a ring
}

\author{{C. Appert-Rolland, J. Cividini, and H.J.~Hilhorst}}

\address{1 - University Paris-Sud; Laboratory of Theoretical Physics\\
Batiment 210, F-91405 ORSAY Cedex, France.}

\address{2 - CNRS; UMR 8627; LPT\\
Batiment 210, F-91405 ORSAY Cedex, France.}

\ead{\mailto{Cecile.Appert-Rolland@th.u-psud.fr}, \mailto{Henk.Hilhorst@th.u-psud.fr}}

\begin{abstract}
\noindent 
We introduce a new rule of motion
for a totally asymmetric exclusion process
(TASEP) representing pedestrian traffic on a lattice. 
Its characteristic feature is that the positions
of the pedestrians, modeled as hard-core
particles, are updated in a fixed predefined order, determined
by a {\em phase} attached to each of them.
We investigate this model analytically and by Monte Carlo
simulation on a
one-dimensional lattice with periodic boundary conditions. 
At a critical value of the particle density  
a transition occurs from a phase with `free flow' to one with 
`jammed flow'.
We are able to analytically predict the current-density diagram
for the infinite system and to find the scaling function
that describes the finite size rounding at the transition point.

\vspace{12mm}
\end{abstract}
\vspace{12mm}

\noindent LPT Orsay 11/37

\maketitle
\newpage

%%%%%%%%%%%%%%%%%%%%%%%%%%%%%%%%%%%%%%%%%%%%%%%%%%%%%%%%%%%%%%%%%%%%%%%%%%%%%
%%%%%%%%%%%%%%%%%%%%%%%%%%%%%%%%%%%%%%%%%%%%%%%%%%%%%%%%%%%%%%%%%%%%%%%%%%%%%

\section{Introduction} 
\label{sect_introduction}

\cappar
Let a set of hard-core
particles, labeled by an index $i=1,2,\ldots,N,$
move unidirectionally from site to site
on a one-dimensional lattice.
We imagine that the particles represent pedestrians all walking at
the same pace but not necessarily in phase with each other.
This leads us to formulate the following rule of motion,
that we state as the update scheme of a Monte Carlo simulation.
Each particle $i$ is assigned a phase $\tau_i\in[0,1)$, permanently
attached to it, and during each time step 
(that is, each unit time interval) 
all $N$ particles make a forward hopping attempt
in the order of increasing phases.
An attempted hop will succeed only if the target site is empty.

This model is an instance of what are commonly called 
Totally Asymmetric Simple Exclusion Processes (TASEP); its novelty 
resides in its update rule. 
Before continuing the discussion, we mention some connections to
existing work.
\vspace{3mm}

Processes of particles moving stochastically
on -- often one-dimensional -- lattices
serve on the one hand
as archetypes of out-of-equilibrium systems, and on the
other hand as modeling tools to study transport in various
systems, ranging from
road and pedestrian traffic to intracellular traffic
\cite{chowdhury_s_n05}. The particle motion may take place according
to a large diversity of hopping rules.
By the `exclusion' principle one imposes the hard core
condition (at most one particle per site);
the `total asymmetry' forbids backward hops; 
and the process is called `simple'
when hops are only between nearest-neighbor sites.

Given these three properties that are characteristic of a TASEP, 
it is still possible to choose from
a variety of update schemes. 
In particular, the following update schemes have been studied: 
parallel update \cite{hinrichsen96,evans97,schadschneider_s98,wolki_s09}, 
random sequential
update, sequential update ordered backward or forward in space
\cite{borodin_f_p07,brankov_p_s04,brankov06,poghosyan_p08,evans97},
sublattice update \cite{fayaz10,pigorsch_s00,poghosyan_p_s10},
and random shuffle update \cite{wolki_s_s06,smith_w07a}.
The properties of the system depend 
on the update scheme \cite{rajewski98} and
the choice of the scheme should be determined by the application.

The most common update schemes are
the random sequential and the parallel updates.
Random sequential update produces a dynamics very
close to that defined by a master equation in continuous
time. A time step is defined as a succession of $N$ elementary
updates, each associated with a time interval of length $1/N$,
and each allowing only a single particle, chosen at random, to make a
hopping attempt. With this dynamics considerable 
fluctuations occur, since the same
particle may be updated several times in the same time step
whereas another one may be ignored during several time steps.

With parallel update particles make hopping attempts
only at integer values of time but then do so simultaneously.
Parallel update is used in particular for applications
to road traffic \cite{nagel_s92,appert_s01}: all vehicles are moving
at the same time and the time step of the scheme is then supposed to represent
a reaction time.
Fluctuations are reduced, but 
parallel update can create conflicts -- 
that should be settled by additional rules --
when more than one particle tries to hop onto the same target site.
This may occur in particular in applications to pedestrian traffic,
which usually takes place in two-dimensional space.

In order to overcome the limitations of these two types of updates,
the so-called `shuffle update' has been proposed for modeling pedestrian
flow.
In the {\it random\,} shuffle update \cite{wolki_s_s06,smith_w07a},
before each time step the
particles are pre-arranged in a randomly chosen order
and then each of them is  updated once, exactly in that order.
This update scheme was used for example in \cite{klupfel07} for
large-scale simulations of pedestrians.
\vspace{3mm}

In the present paper we explore a variant of the shuffle update
for which the order in which the particles are arranged, 
that is, the updating order, is fixed once and for all%
\footnote{This frozen variant was mentioned, but not studied,
in the conclusion of \cite{wolki_s_s06}.}.
Our scheme is therefore appropriately characterized by the name of
`{\it frozen\,} shuffle update'.

For a {\it closed\,} system this is easy to implement; 
a random phase $\tau_i$ is drawn for each of the $N$ particles
independently, for example from the uniform distribution on $[0,1)$.
In each time step all $N$ particle positions are updated once, 
one after the other,
{\it according to increasing values of their phases}.
The phases $\tau_i$ do not change during the whole simulation
and may be considered as frozen variables of the motion.
The set $\{\tau_i\}$ 
determines a random permutation of the particles; for uniformly distributed
$\tau_i\,$, all permutations have the same probability.

A closed system is expected to evolve towards a stationary state.
We must be prepared to envisage that
the final stationary state may depend
(and as we shall see, indeed does depend) 
on the precise permutation that fixes the updating order of the particles.
An average over all permutations is
therefore appropriate and is
analogous to the averages on quenched disorder
variables standardly performed in statistical physics.
The term `disorder average' will therefore denote below
the average over all random assignments $\{\tau_i\}$.

For an open system
the frozen shuffle update requires that by a suitable algorithm
we fix the phase of each particle the moment it enters. 
The equivalence of the set
$\{\tau_i\}$ to a simple permutation may then no longer hold.
The case of open boundary conditions will not be considered here
but is studied in a forthcoming paper \cite{appert-rolland_c_h11b}.
\vspace{3mm}

In section \ref{sect_defs} we introduce
some terminology that actually already is the expression
of several model properties.
In section \ref{sect_pbc} we consider the TASEP 
with frozen shuffle update 
on a ring with particle density $\rho$. 
We show that a phase transition occurs
at a critical density $\rho_{\rm c}$
which separates a low density regime with `free flow' 
from a high density regime with `jammed flow'.
We determine the current-{\it versus}-density curve $J_L(\rho)$
analytically,
first for an infinite system
(section \ref{sect_pbcinfinite}) where 
$J(\rho)=\lim_{L\to\infty}J_L(\rho)$ 
has a cusp, and then for a system of finite size
$L$ (section \ref{sect_pbcfinite}), where the finite size rounding 
of $J_L(\rho)$ is described by a scaling function that depends only on the
product variable $(\rho-\rho_{\rm c})L^{1/2}$.
Monte Carlo simulations show very good agreement with theory.
In section \ref{sect_continuous}, by way of a supplement,
we show that under `free flow' conditions the TASEP
with frozen shuffle update is equivalent
to a system of noninteracting particles in
continuous space and time.
In section \ref{sect_conclusion} we conclude.

%%%%%%%%%%%%%%%%%%%%%%%%%%%%%%%%%%%%%%%%%%%%%%%%%%%%%%%%%%%%%%%%%%%%%%%%%%%%

\section{Free flow and jammed flow}
\label{sect_defs}
We introduce here the concepts that characterize 
the structures formed by
the particles as they result from the frozen
shuffle update scheme.
The discussion below
is independent of any boundary conditions that may be imposed at a
later stage.
The most important point is the identification of 
two distinct stable flow states
that we call the {\it free flow state\,} and the {\it jammed state.}

%%%%%%%%%%%%%%%%%%%%%%%%%%%%%%%%%%%%%%%%%%%%%%%%%%%%%%%%%%%%%%%%%%%%%%%%%%%%%

\subsection{Well-ordered  and ill-ordered pairs}
\label{sect_well_ill}

\begin{figure}
\begin{center}
\scalebox{.40}
{\includegraphics{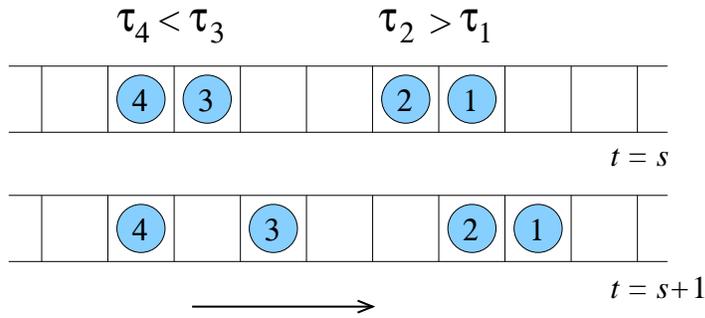}}
\end{center}
\caption{\small
Lattice sites are represented by squares that may be
either empty or occupied by a single particle.
A configuration involving four particles
is shown at two successive times $t=s$ and $t=s+1$;
the flow is in the direction of the arrow.
Particles 1 and 2 form a
well-ordered pair. They can move at the same
time step, as particle 1 is updated before
particle 2.
Particles 3 and 4 form an ill-ordered pair.
During the time step from $t=s$ to $t=s+1$
the update attempt of particle 4 is performed at time $s+\tau_4$
but remains unsuccessful,
since its target site is still occupied
by particle 3. When subsequently particle
3 is updated at time $s+\tau_3$, it moves forward.
Thus a hole is inserted between the two particles.
In case $\tau_3 < \tau_2$, particles 2 and 3 also form
an ill-ordered pair, but they are not
adjacent and thus can hop independently.
}
\label{fig_pairs}
\end{figure}

Let the flow direction be to the right,
let the particles be numbered $\ldots,i-1,i,i+1,\ldots$
from right to left (see figure \ref{fig_pairs}),
and let their phases 
$\ldots,\tau_{i-1},\tau_i,\tau_{i+1},\ldots$ be given.
The pair of successive
particles $(i,i+1)$, not necessarily on adjacent sites, 
will be called {\it well-ordered\,} if $\tau_{i+1} > \tau_i$
and {\it ill-ordered\,} in the opposite case.
The time evolution of well- and ill-ordered
pairs under the frozen shuffle update scheme has the following properties,
illustrated in figure \ref{fig_pairs}.

%%%%%%%%%%%%%%%%%%%%%%%%%%%%%%%%
%%%%%%%%%%%%%%%%%%%%%%%%%%%%%%%%
%%%%%%%%%%%%%%%%%%%%%%%%%%%%%%%%
%%%%%%%%%%%%%%%%%%%%%%%%%%%%%%%%

If a well-ordered pair $(i,i+1)$ occupies two adjacent sites, 
then at each time step particle $i$ will
move first and particle $i+1$ will move next; hence when the time step
is completed, the two particles are still adjacent and have advanced
one lattice distance to the right. Their speed is $v=1$ in units of
lattice distances per time step.

If an ill-ordered pair occupies two adjacent sites, the two particles
cannot move in the same time step; particle $i+1$, 
having $\tau_{i+1}<\tau_i$, will attempt first to move but finds
itself blocked by particle $i$. Hence the two particles of
an ill-ordered pair move at speed $v=1$ 
only if they are separated by at least one empty site.

\subsection{Free flow configuration}
\label{sect_freeflow}

A particle configuration will be said to
satisfy the {\it free flow\,} (FF) 
condition
when {\em each} ill-ordered pair
has its two members separated by at least one hole.
In view of the above,
such a configuration is again identical to itself
at the end of each time step
except for a translation by one lattice distance to the right. 
This corresponds to a free flow with speed $v=1$, and 
hence for a FF configuration we have
\beq
J=\rho,
\label{xJff}
\eeq
where $J$ is the current and $\rho$ the particle density.
It is tacitly understood here that these quantities refer to
time averages in a stationary state.

%%%%%%%%%%%%%%%%%%%%%%%%%%%%%%%%%%%%%%%%%%%%%%%%%%%%%%%%%%%%%%%%%%%%%%%%%%%%%

\subsection{Rising sequences and platoons}
\label{sect_platoon}

\begin{figure}
\begin{center}
\scalebox{.40}
{\includegraphics{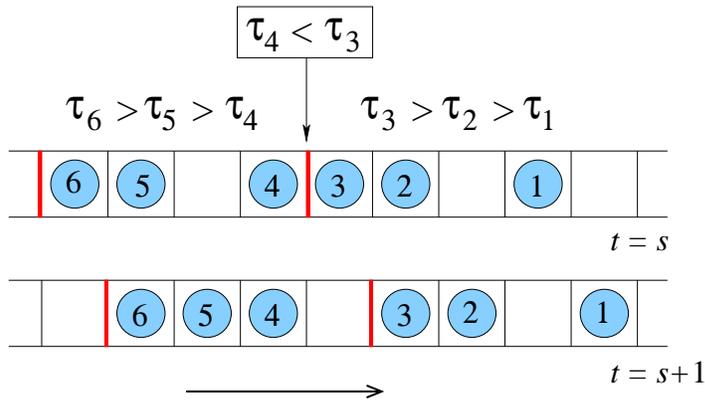}}
\end{center}
\caption{\small
  A six particle configuration is shown at two successive 
  times $t=s$ and $t=s+1$.
  Because of the inequalities between their phases,
  the set of particles $(1,2,3)$ forms a rising
  sequence, and so does $(4,5,6)$.
  The inequality
  $\tau_3>\tau_4$ defines the separation between
  these two sequences. The fact that
  particles $3$ and $6$ are the last ones of their rising sequences is
  marked by a heavy (red) line segment delimiting their lattice site
  to the left. 
  At the $(s+1)$th time step all particles will move except 
  number $4$. 
  It so happens that
  at time $t=s+1$ particles $4$, $5$, and $6$ have formed a platoon,
  {\it i.e.} the rising sequence $(4,5,6)$ has been compacted. 
}
\label{fig_flow}
\end{figure}

Going along the lattice from right to left one may divide the
particles encountered into sequences of
increasing phases (for short: {\it rising sequences}).
The set of particles $(i,i+1,\ldots,i')$ 
will be said to constitute a rising sequence if 
$\tau_{i} < \tau_{i+1} < \ldots <\tau_{i'}$
but $\tau_{i-1} > \tau_{i}$ and $\tau_{i'} > \tau_{i'+1}$.
Examples are shown in figure \ref{fig_flow}.

%%%%%%%%%%%%%%%%%%%%%%%%%%%%%%%%
%%%%%%%%%%%%%%%%%%%%%%%%%%%%%%%%
%%%%%%%%%%%%%%%%%%%%%%%%%%%%%%%%
%%%%%%%%%%%%%%%%%%%%%%%%%%%%%%%%

Let a set of particles 
occupy consecutive sites and have phases that
increase from right to left.
If this set corresponds to a full rising sequence, it will
be called a {\em platoon}.
If it corresponds to only part of a rising sequence,
it will be called a {\em subplatoon}.
One may say that a platoon (a subplatoon)
is a fully compacted rising sequence (part of a rising sequence).
Platoons and subplatoons are limited on both ends either by holes or
by ill-ordered pairs.
A rising sequence is composed either
of a single platoon or of several subplatoons, 
an isolated particle being considered as a (sub-)platoon of length $1$.

Under the frozen shuffle update
platoons and subplatoons obey the following simple rules.

${}$\phantom{i}(i) If in a given time step the first particle of a
(sub-)platoon can move, then all its other particles will also move;
hence (sub-)platoons move as single entities. 

(ii) When two subplatoons merge, they can
never separate again; hence the length of a subplatoon can only grow
until it includes the whole rising sequence in which
it is embedded.

%%%%%%%%%%%%%%%%%%%%%%%%%%%%%%%%%%%%%%%%%%%%%%%%%%%%%%%%%%%%%%%%%%%%%%%%%%%%%%

\subsection{Jammed configuration}
\label{sect_jammed}

%%%%%%%%%%%%%%%%%%%%%%%%%%%%%%%%
%%%%%%%%%%%%%%%%%%%%%%%%%%%%%%%%
\begin{figure}
\begin{center}
\scalebox{.40}
{\includegraphics{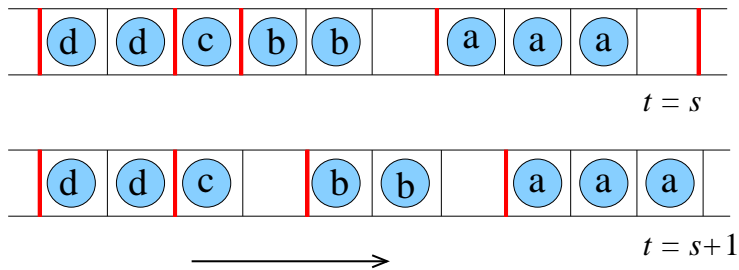}}
\end{center}
\caption{\small A jammed configuration involving seven particles is
  shown at two successive
  time steps $t=s$ and $t=s+1$. All particles are grouped together in
  platoons; particles belonging to the same platoon are
  labeled by the same letter. The last particle of each platoon is
  indicated by a heavy (red) line segment to its left. 
  Successive platoons are separated by either zero or a single
  hole, that is, the particles are in a jammed configuration. 
  During the $(s+1)$th time step platoons $a$ and $b$ move one lattice 
  distance to the right, but $c$ and $d$ are blocked. Inversely, one
  may describe this dynamics as a motion of holes that jump at
  each time step across the platoon to their left.
}
\label{fig_jam}
\end{figure}
%%%%%%%%%%%%%%%%%%%%%%%%%%%%%%%%
%%%%%%%%%%%%%%%%%%%%%%%%%%%%%%%%

A configuration of particles will be called {\it jammed}
if all its rising sequences are compacted into platoons
and if consecutive platoons are
separated by at most one hole.
Figure \ref{fig_jam} shows 
an example of a jammed configuration.
At each time step the evolution of a jammed configuration
may be simply described in terms of the motion of its platoons.
The rules follow directly from those above:

${}$\phantom{i}(i) A platoon preceded by a hole
advances by one site as a single entity; 
this amounts to a position exchange of the platoon and the hole.

(ii) A platoon not preceded by a hole is blocked and does not advance.

Let $\nu$ stand for the average platoon length in 
a jammed configuration that is statistically homogeneous in space.
Noting that
$1-\rho$ is the hole density and that only
platoons preceded by a hole move,
we may write the particle current in this jammed configuration as
\beq
J=(1-\rho)\nu.
\label{xJjam}
\eeq
Obviously, configuration space has many configurations that are
neither `free flow' nor `jammed' in the sense of the above definitions.
There is also one subclass of configurations that are both `free flow' and
`jammed'; this happens when {\it all\,} platoons of a jammed
configuration are separated by exactly one hole.

%%%%%%%%%%%%%%%%%%%%%%%%%%%%%%%%%%%%%%%%%%%%%%%%%%%%%%%%%%%%%%%%%%%%%%%%%%%%%

\section{Phase transition on a ring}
\label{sect_pbc}

After the preliminaries of section \ref{sect_defs}
we are now ready to study a concrete system.
We consider a ring, that is,
a lattice of $L$ sites with periodic boundary conditions.
Let $N$ be the number of particles and hence
$\rho = N/L$ the particle density.
We set ourselves the purpose of determining
the particle current $J_L(\rho)$
as a function of the particle density $\rho$ in the stationary state
that will result from a given initial state.

At the initial time $t=0$ the particles are placed at distinct but otherwise
random positions on the lattice. They are numbered
$i=1,2,\ldots,N$ from right to left (clockwise around the lattice)
and their direction of motion is from left to right (anticlockwise).
The particles are assigned
a random phases $\tau_i$
which we take independently and uniformly distributed on $[0,1)$.
This assignment determines the updating order of the particles.

The initial
configuration does not necessarily satisfy the FF condition.
If it does, then the particle configuration at time $t=s$ is obtained
from that at time $t=0$ by rotating all particle
positions by $s$ steps along the ring. 
If the FF condition is not satisfied, then  
after a transient period the system will reach a stationary
state which may or may not be of the FF type.
We will investigate below
the conditions for the realization of each of these possibilities,
and the ensuing consequences for the particle current.

%%%%%%%%%%%%%%%%%%%%%%%%%%%%%%%%%%%%%%%%%%%%%%%%%%%%%%%%%%%%%%%%%%%%%%%%%%%%

\subsection{Infinite system limit}
\label{sect_pbcinfinite}

The infinite system limit is easiest to discuss,
since we may apply the law of large numbers and formulate
statements that in that limit are true with probability 1. 
Let us first ask up to which value of the density $\rho$
it is still possible to have free flow.

For a given set $\{\tau_i\}$ the densest possible FF
configuration occurs when all rising sequences 
are compacted into platoons separated by a single hole. 
This corresponds precisely to the aforementioned
special case of a configuration which is both FF and jammed.
The maximum density of the FF phase thus is 
$\rho_{\rm c}=\nu/(\nu+1)$,
where as before $\nu$ is the average platoon length.
It may be shown (see \cite{oshanin_v04} or Appendix \ref{sect_permutation}) 
that in the infinite system limit
one has $\nu=2$ when the phases $\tau_i$ are uniformly distributed,
and therefore $\rho_{\rm c}=\frac{2}{3}$\,. 

For $\rho\leq\rho_{\rm c}$ any arbitrary initial configuration
-- tacitly understood to be statistically homogeneous in space --
will, after a transient, be converted into a FF configuration.
Indeed, whenever an ill-ordered pair of particles occupies two
successive sites,
the second one will not yet be able to move
when the first one first moves forward, 
and a hole will naturally be included between them.
When in this way all ill-ordered pairs have come to include
a hole, a FF configuration is obtained.
The current $J(\rho)=\lim_{L\to\infty}J_L(\rho)$ 
is then given by its FF value (\ref{xJff}),
\beq
J(\rho)=\rho, \qquad \rho \leq \rho_{\rm c}\,.
\label{xJavff}
\eeq

\begin{figure}
\begin{center}
\scalebox{.45}
{\includegraphics{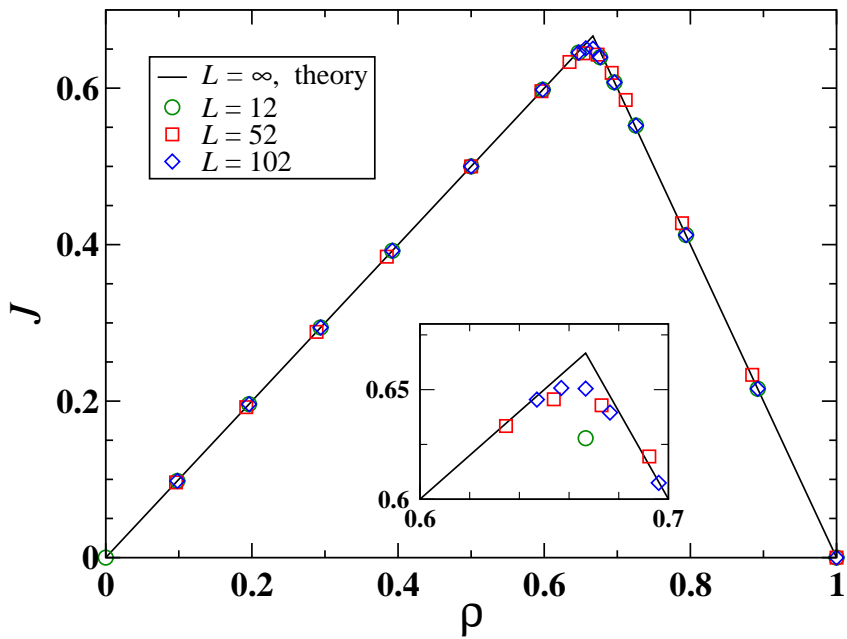}}
\end{center}
\caption{\small 
Current ${J}$ as a function
of the density $\rho$ for periodic boundary conditions.
Solid line: theoretical prediction for an infinite system.
Data points: Monte Carlo
simulations for systems of size $L=12$, $52$, and $102$.
The inset is a zoom around the maximum of the curve.
}
\label{fig_Linftypbc}
\end{figure}

For $\rho > \rho_{\rm c}$ the time evolution will
produce two effects.
It will compact rising sequences into
platoons and it will distribute the available holes
such that each platoon is separated from its predecessor by at most a
single hole. However, the number of holes
is less than the number of platoons.
The number of platoons that move in a given time
step has thus been maximized and is equal to the number of holes, 
the other platoons being blocked at that time step. 
This corresponds to the definition of a jammed phase given
in section \ref{sect_jammed}, whence
upon applying (\ref{xJjam}) with $\nu=2$ we obtain
\beq
J(\rho) = 2(1-\rho), \qquad \rho\geq\rho_{\rm c}\,.
\label{xJavjam}
\eeq
Equations (\ref{xJavff}) and (\ref{xJavjam})
lead to the cusped current-density diagram shown in
figure \ref{fig_Linftypbc}.
The agreement with finite size Monte Carlo simulations
is already quite good for system size $L=12$.
However, finite size effects are 
visible around the maximum, as shown in the inset
of figure \ref{fig_Linftypbc}.
In the next section we shall refine the theory to account
for this rounding of the transition.

%%%%%%%%%%%%%%%%%%%%%%%%%%%%%%%%
%%%%%%%%%%%%%%%%%%%%%%%%%%%%%%%%
%%%%%%%%%%%%%%%%%%%%%%%%%%%%%%%%
%%%%%%%%%%%%%%%%%%%%%%%%%%%%%%%%

%%%%%%%%%%%%%%%%%%%%%%%%%%%%%%%%%%%%%%%%%%%%%%%%%%%%%%%%%%%%%%%%%%%%%%%%%%%%

\subsection{Finite system}
\label{sect_pbcfinite}

We consider in this subsection
a finite ring of size $L$ containing
exactly $N$ particles; throughout we set $\rho=N/L$.
Our interest is in the density dependent particle current, which in
this finite system we shall denote by $J_L(\rho)$ and whose
definition we shall render precise. 

By the mechanism described above the system, whatever its initial
configuration, will evolve so as to maximize
the number of ill-ordered pairs that include a hole.
For densities $\rho \leq 1/2$, 
there is enough space in the system to place a hole
between {\em each\,} pair of particles. 
Then the FF condition can be fulfilled with certainty
and the stationary state is a FF state.
Denoting the current
in the stationary state
by $J_{NL}$ we have
\beq
J_{NL} = \frac{N}{L} = \rho,  \qquad \rho\leq\tfrac{1}{2}. 
\label{xJavff_finite}
\eeq
For densities $\rho>1/2$ it may or may not be possible
to converge towards an FF configuration, depending
on the random assignment $\{\tau_i\}$.
The considerations of section \ref{sect_freeflow} show that
the discriminating quantity is the
number of ill-ordered particle pairs in the initial state.
We denote by $\ok$ (by $\pok$) the number of well-ordered
(ill-ordered) pairs, so that $\ok+\pok=N$.
It will be convenient to work with the difference variable
\beq
\dok\big(\{\tau_i\}\big)=\ok-\pok\,,             
\label{defdok}
\eeq
of which we shall henceforth suppress the argument.
Because of the periodic boundary conditions, there is always at least
one ill-ordered pair and one well-ordered pair
in the system, so that $\dok$ may take the values
$\dok=1,2,\ldots,N-1$.
A necessary and sufficient condition to fulfill the FF condition in the
stationary state is to have at least
one empty site available for each ill-ordered pair, that is,
$\ok + 2 \pok \le L$
or equivalently
\beq
\dok \geq \tfrac{3}{2}N-L.
\label{dok_condition}
\eeq
The expression for the
stationary state current now involves the variable $n$ and we will
denote it by $J_{NLn}$. Two cases have to be distinguished.
First, if inequality (\ref{dok_condition}) is satisfied, the
system evolves towards a FF state
and for this subset
of realizations the current is
\begin{equation}
J_{NLn}=\frac{N}{L}=\rho, \qquad \rho>\tfrac{1}{2}\,, \quad
\frac{n}{N} \geq \tfrac{3}{2}-\rho^{-1}\,. 
\label{j_ff}
\end{equation}
Secondly, we consider realizations $\{\tau_i\}$ for which
inequality (\ref{dok_condition}) is violated. 
The stationary state then only has isolated holes%
\footnote{The same behavior appears with random shuffle update.}, 
and all rising sequences are compacted into platoons.
At each time step only the platoons headed by
one of the $L-N$ holes move forward, which means
that the instantaneous current per time step fluctuates with time.
However, averaged over time 
each platoon will
move in a fraction $(L-N)/N$ of all time steps.
Using the fact that
${N}/{\pok}=2/(1-\dok/N)$ is the
average length of a platoon, 
we therefore find after time averaging for the current
$J_{NLn}$ the expression
\begin{equation}
J_{NLn}= \frac{L-N}{L}\times\frac{N}{\pok}
= \frac{2(1-N/L)}{1-\dok/N}, \qquad \rho>\tfrac{1}{2}\,, \quad
\frac{\dok}{N} < \tfrac{3}{2}-\rho^{-1}.
\label{j_jam}
\end{equation}
For finite systems and for densities $\frac{1}{2}<\rho<1$,
there will always exist realizations of the $\tau_i$ that converge
towards FF stationary states with a current $\rho$,
and others that do not 
and have a current less than $\rho$ and given by (\ref{j_jam}).
In this density regime we will denote by $J_{NL}$ 
the current $J_{NLn}$ of (\ref{j_ff}) and (\ref{j_jam}) {\it averaged\,}
with respect to $\dok$,
that is,
\beq
J_{NL}= \sum_{\dok=1}^{N-1}P_N(\dok) J_{NLn}\,, \qquad \rho>\tfrac{1}{2}\,,
\label{Jav}
\eeq
in which $P_N(\dok)$ is 
the probability distribution of $\dok\big(\{\tau_i\}\big)$ 
and remains to be determined.
Since $n$ is determined by $\{\tau_i\}$, the current $J_{NL}$ in
(\ref{Jav}) deserves the name of `disorder averaged current'.

%%%%%%%%%%%%%%%%%%%%%%%%%%%%%%%%%%%%%%%%%%%%%%%%%%%%%%%%%%%%%%%%%%%%%%%%%%%%

\subsection{Finite size effects near the transition point}
\label{sect_pbcrounding}

The probability distribution $P_N(n)$ was studied by Oshanin and
Voituriez \cite{oshanin_v04} for the case -- which is also ours --
where the $\tau_i$ are
drawn independently from a uniform distribution on $[0,1)$. 
These authors showed, among other things,
that in the limit of large $N$ and with $n$ scaling as $\sim N^{1/2}$
the variable $x=\dok/N^{1/2}$
has the probability distribution
\beq
\Pi(x) =  (3/2\pi)^{1/2}\exp\left( -\tfrac{3}{2}x^2\right).
\label{PiNx}
\eeq
It is symmetric in $n$, as dictated by the left-right symmetry of the 
phase assignment.
In Appendix \ref{sect_permutation} we derive equation (\ref{PiNx}) 
in a more direct way.

\begin{figure}
\begin{center}
\scalebox{.45}
{\includegraphics{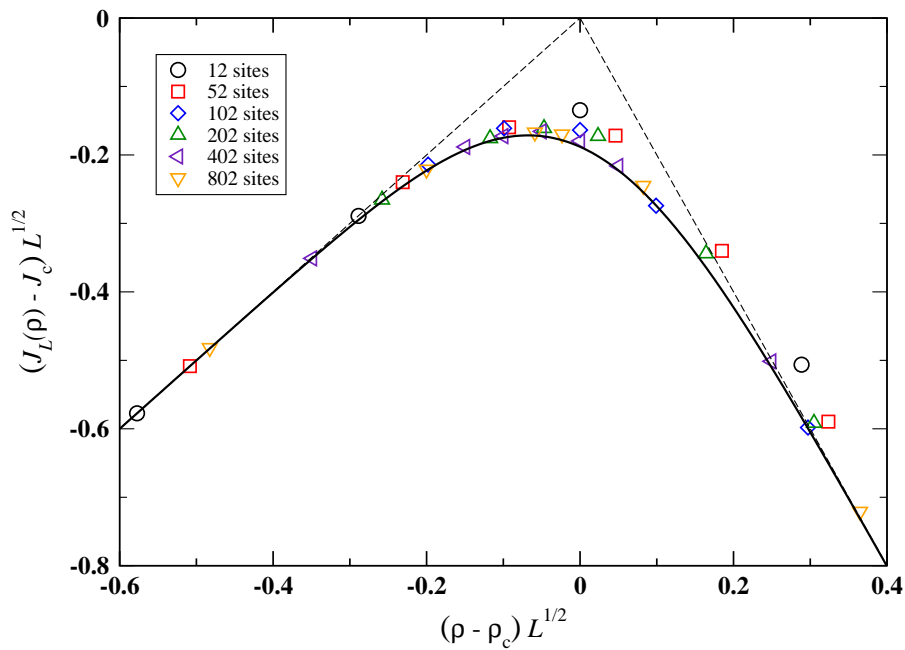}}
\end{center}
\caption{\small 
Solid line: the theoretical
scaling function $\Phi\big((\rho-\rho_{\rm c}\big)L^{1/2})$ 
of equation (\ref{xPhiy}),
representing the
average current $J_L(\rho)-J_{\rm c}$ as a function
of the particle density $\rho$ in a finite system
of size $L$ near criticality.
The dashed lines are the asymptotes  for 
$(\rho-\rho_{\rm c})L^{1/2}\to\pm\infty$.
Simulation data  
for large system sizes $L$ are seen to collapse very well 
onto the theoretical curve.
Each point corresponds to an average over $10\,000$ or $100\,000$
time steps and over $1000$ realizations of the disorder.
}
\label{fig_scaling}
\end{figure}

From here on we shall consider the equations of the preceding
subsection in the limit of large but finite $N$, $n$ and $L$, 
and fixed ratios $\rho=N/L$ and $x=n/N^{1/2}$. 
We will conform to usage and
take the system size $L$, rather than $N$, as the independent large variable.
In the limit in question we shall write 
$J_{NLn} = J_L(\rho,x)$ and $J_{NL}=J_L(\rho)$.
We may then reexpress the disorder averaged current (\ref{Jav}) as 
\beq
J_L(\rho) = \int_{-\infty}^{\infty}\!{\rm d}x\,\Pi(x)J_L(\rho,x).
\label{jav}
\eeq
The expression for $J_L(\rho,x)$ is derived from (\ref{j_ff}) or (\ref{j_jam}),
depending on the value of $x$, that is,
\beq
J_L(\rho,x)=\left\{
\begin{array}{ll}
\frac{2(1-\rho)}{1-x(\rho L)^{-1/2}},\qquad & x<x_{\rm c}(\rho), \\[2mm]
\rho, & x \geq x_{\rm c}(\rho),
\end{array}
\right.
\label{xjx}
\eeq
in which
\beq
x_{\rm c}(\rho) = (\rho L)^{1/2}\left(\rho_{\rm c}^{-1}-{\rho}^{-1}\right)
\label{defxc}
\eeq
where $\rho_{\rm c}=\frac{2}{3}$.
We observe parenthetically
that in the limit $L\to\infty$ the $x$ dependence of
(\ref{defxc}) disappears and we recover 
$\lim_{L\to\infty}J_L(x,\rho)=J(\rho)$, where $J(\rho)$ is the
infinite system current of 
equations (\ref{xJavff}) and (\ref{xJavjam}).
Since (\ref{PiNx}) is valid in the limit in which $x$ remains finite
as $L\to\infty$, we conclude that the present approach is valid for
densities
\beq
\rho=\rho_{\rm c}+\Delta\rho
\label{defDeltarho}
\eeq
such that $\Delta\rho$ is on the scale of $L^{-1/2}$.
Remembering this and expanding in powers of $L^{-1/2}$ we find
from (\ref{defxc}) and (\ref{xjx})
\bea
x_{\rm c}(\rho) &=& (\tfrac{3}{2})^{3/2}L^{1/2}\Delta\rho 
                    + {\cal O}(L^{-1/2}),
\nonumber\\[2mm]
J_L(\rho,x)      &=& \left\{
\begin{array}{ll}
\rho_{\rm c} - 2\Delta\rho + (3L/2)^{-1/2}x 
+ {\cal O}(L^{-1}),        & x   <  x_{\rm c}(\rho),\\[2mm]
\rho_{\rm c} + \Delta\rho, & x \geq x_{\rm c}(\rho).
\end{array}
\right.
\label{expansion1}
\eea
We introduce the scaling variable $y = L^{1/2}\Delta\rho$, which in
the limit of interest should be of order unity.
Substitution of (\ref{expansion1}) in (\ref{jav}) then yields
\beq
J_L(\rho) = \rho_{\rm c} + L^{-1/2} y - 
L^{-1/2}\int_{-\infty}^{(3/2)^{3/2}y}\!{\rm d}x\,\Pi(x)
\left[ 3y - (\tfrac{2}{3})^{1/2}x \right] + {\cal O}(L^{-1}).
\label{resjav2}
\eeq
When using in (\ref{resjav2})
the explicit expression (\ref{PiNx}) for $\Pi(x)$  we may evaluate the 
$x$ integral and obtain,
up to corrections of higher order in $L^{-1/2}$,
\beq
J_L(\rho) = \rho_{\rm c} + L^{-1/2}\Phi(L^{1/2}\Delta\rho),
\label{resfinaljav}
\eeq
valid in the limits $\Delta\rho=\rho-\rho_{\rm c}\to 0$ and $L\to\infty$
with $L^{1/2}\Delta\rho$ fixed,
and in which the scaling function $\Phi(y)$ is given by
\beq
\Phi(y) = -\tfrac{1}{2}y
- \tfrac{3}{2}y\,{\rm erf}\left( \tfrac{9}{2}y \right)
- (9\pi)^{-1/2}
\exp\left( -\tfrac{81}{4}y^2 \right).
\label{xPhiy}
\eeq
This function is negative and such that
\bea
\Phi(y) &\simeq&   y,\phantom{\Delta\rho} \qquad y\to -\infty, \nonumber\\[2mm]
\Phi(y) &\simeq& -2y,           \qquad y\to  \infty, 
\label{propPhiy}
\eea
which ensures the correct limit
behavior of (\ref{resfinaljav}) for $|\rho-\rho_{\rm c}|\gg L^{-1/2}$.
We have plotted $\Phi(y)$ in figure \ref{fig_scaling}
together with simulation data for different system sizes $L$.
The data are seen to collapse very well onto the theoretical curve.

%%%%%%%%%%%%%%%%%%%%%%%%%%%%%%%%
%%%%%%%%%%%%%%%%%%%%%%%%%%%%%%%%
%%%%%%%%%%%%%%%%%%%%%%%%%%%%%%%%
%%%%%%%%%%%%%%%%%%%%%%%%%%%%%%%%

%%%%%%%%%%%%%%%%%%%%%%%%%%%%%%%%%%%%%%%%%%%%%%%%%%%%%%%%%%%%%%%%%%%%%%%%%%%%%

\section{Mapping to a continuous model and 
interpretation for pedestrian motion}
\label{sect_continuous}

In this section we point out 
that under free flow conditions the time evolution 
defined by the frozen shuffle update for the particle system on a lattice
may be seen as a sequence of snapshots
taken at integer instants of time $t=\ldots,s-1,s,s+1,\ldots$, 
of a system that itself evolves in 
continuous time $t$ and space $x$.

\begin{figure}
\begin{center}
\scalebox{.50}
{\includegraphics{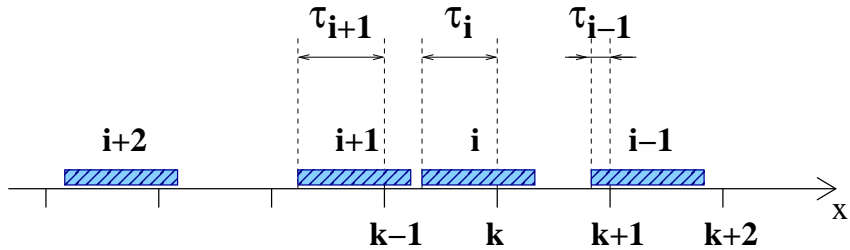}}
\end{center}
\caption{\small Hard rods labeled by an index $i$ move at constant
speed $v=1$ along the $x$ axis. The integer axis positions have been labeled
by an index $k$.
}
\label{fig_mapping}
\end{figure}

To show this
we consider a collection of nonoverlapping
hard rods all moving continuously to the right at speed $v=1$ along
the $x$ axis,
as depicted in figure \ref{fig_mapping}.
If we associate lattice sites with the integer axis positions
$x=\ldots,k-1,k,k+1,\ldots$,
then at any given instant of continuous time,
each rod covers exactly one site.
The mapping is performed by placing on that site a particle associated with
that rod. Let figure \ref{fig_mapping} represent the rod positions at
time $t=0$ (or for that matter at any other integer instant of time).
The particle labeled $i$ and corresponding to rod $i$
occupies lattice site $k$ and therefore gives rise, at $t=0$, 
to a particle on site $k$. 
The distance between site $k$
and the tail of rod $i$
has been indicated as a time interval% 
\footnote{Because $v=1$, times and distances may be identified.}
$\tau_i$, this being the time still needed for the tail of that rod 
to cross the point $k$ during its continuous motion along the $x$
axis. This crossing therefore occurs at time $t=\tau_i$, and that is
the time at which particle $i$ will hop from site $k$ to site
$k+1$. Particle $i$ will execute its subsequent hops at times
$t=s+\tau_i$, where $s$ is an integer.
This is exactly the frozen shuffle update scheme.

We remark that the mapping defined here yields only the FF
configurations of the discrete model.
If we try to perform the inverse mapping, {\it i.e.} from the discrete
to the continuous model, 
then in case of a jammed configuration the non-overlapping
condition for rods cannot be enforced anymore.
This may actually still have some physical relevance, if
one adopts the view that a rod represents not only a pedestrian
but also some ``private'' space around him.
In free flow pedestrians are
not willing to approach each other too closely and they avoid to
enter each other's ``private'' space, whereas
at increasing densities they tolerate smaller distances.

%%%%%%%%%%%%%%%%%%%%%%%%%%%%%%%%
%%%%%%%%%%%%%%%%%%%%%%%%%%%%%%%%
%%%%%%%%%%%%%%%%%%%%%%%%%%%%%%%%
%%%%%%%%%%%%%%%%%%%%%%%%%%%%%%%%

%%%%%%%%%%%%%%%%%%%%%%%%%%%%%%%%%%%%%%%%%%%%%%%%%%%%%%%%%%%%%%%%%%%%%%%%%%%%%

\section{Conclusion}
\label{sect_conclusion}

We have introduced in this paper a new update scheme
for the TASEP, namely the frozen shuffle update,
which should be appropriate, in particular, for
the modeling of pedestrians.

We have characterized the behavior of the TASEP
with frozen shuffle update for a closed one-dimensional lattice of
$L$ sites and $N$ particles. The time evolution
under frozen shuffle update is deterministic%
\footnote{By this we mean that the hopping probability is always unity
when the target site is empty.}; it is
fully determined
by the initial particle positions
and by the set $\{\tau_i\}$ of their phases. The latter are
quenched random variables that at each time step determine the
update order of the particles.
We showed that
the analysis of the particle motion and their interaction
may be fruitfully carried out in terms of the concepts of
well/ill-ordered pairs and of platoons.
Two principal types of flow may then be
distinguished, `free flow' and `jammed flow'.

We were able to predict completely
the fundamental diagram, that is, the current $J_L(\rho)$ as a function
of density $\rho=N/L$, both for the infinite ($N,L\to\infty$)
and the finite system.

We found that
for increasing particle density $\rho$ the passage from
a free flow phase to a jammed phase takes place {\it via} 
a phase transition
at a critical density $\rho=\rho_{\rm c}$.
This contrasts with the random sequential
update, for which with increasing density
the system becomes gradually more and more congested.
Critical points, however, were observed in the fundamental
diagram in other instances of deterministic motion, namely
with parallel update \cite{schadschneider_s93,eisenblatter98}
and with random shuffle update \cite{wolki_s_s06,smith_w07a}.
In the latter case, although the particle-hole
symmetry is broken, the critical point was still found
at the symmetric point $\rho_c=1/2$;  by contrast, for the present frozen
shuffle update we find $\rho_c=2/3$, {\it i.e.} the asymmetry
between holes and particles is still enhanced.
Another difference is that for the random shuffle
update the critical point is already present in
finite systems, whereas for the frozen shuffle update
the transition is rounded and becomes sharp only in
the limit of infinite system size.

A mapping with a continuous model of hard rods is proposed,
which is exact for free flow configurations, and 
may be useful for the interpretation of the results
in terms of pedestrian motion.

Two final remarks about open questions are in place here.
First, the deterministic time evolution studied in this paper entails that,
if the target site is empty, particles hop with probability $p=1$.
Whereas in the case of a random sequential update
the hopping probability $p$ can be modified through a simple
rescaling of time, here such a rescaling is not possible.
We therefore expect a qualitatively different behavior of the
current $J(\rho)$ when the hopping
probability $p$ is strictly less than one.
We leave the analysis of this case
for future work.

Second, this work has been exclusively concerned with a closed system.
New types of 
questions arise when one applies frozen shuffle update to open systems.
In a companion paper \cite{appert-rolland_c_h11b}
we shall address the 
case of open boundary conditions and determine in particular
the phase diagram.

%%%%%%%%%%%%%%%%%%%%%%%%%%%%%%%%%%%%%%%%%%%%%%%%%%%%%%%%%%%%%%%%%%%%%%%%%%%%%%

%\section*{Acknowledgments}

%The authors thank \ldots

%%%%%%%%%%%%%%%%%%%%%%%%%%%%%%%%%%%%%%%%%%%%%%%%%%%%%%%%%%%%%%%%%%%%%%%%%%%%%%

\appendix 

\section{Random walk generated by a random 
permutation of $N$ integers}
\label{sect_permutation}

We arrange the integers $1,2,3,\ldots,N$ on the sites of a circular
lattice and permute them randomly, all permutations having the same
probability.
Suppose that when
going clockwise
along the lattice in $N$ steps, we encounter $\ok$
well-ordered and $\pok$ ill-ordered pairs
in the sense of section \ref{sect_well_ill}. Obviously $\ok$ and $\pok$ are
random integers that depend on the permutation, and are such that
$\ok+\pok=N$.
Let $\dok=\ok-\pok$. We ask what the probability distribution $P_N(\dok)$ of
$\dok$ is in the limit of large $N$.

This question was first asked by Oshanin and Voituriez
\cite{oshanin_v04},
who obtained the distribution $\Pi(x)$ given in (\ref{PiNx}). 
It is possible to arrive at same result in a different and, we
believe, simpler way that we present here. It is based on
establishing a recursion in $N$.
Suppose that the integers $1,2,\ldots,N$
have been permuted and placed on the sites of a circular $N$-site lattice.
A permutation of $1,2,\ldots,N+1$ on an $(N+1)$-site lattice
is obtained by inserting between two
randomly chosen neighboring sites a new site carrying the integer $N+1$.
The probability $p_N^{\rm w}$ (or $p_N^{\rm i}$)
to perform the insertion on a well-ordered (or on an ill-ordered) pair is
\beq
p_N^{\rm w,i}(\dok)=\tfrac{1}{2}[1\pm \dok/N].
\label{pNupdown}
\eeq
In either case the original pair disappears and, since the newly
inserted integer $N+1$ is necessarily larger than its two neighbors,
is replaced with the succession of a well- and an ill-ordered pair.
Hence we have the recursion
\beq
P_{N+1}(\dok) = p_N^{\rm w}(\dok-1)P_N(\dok-1) 
              + p_N^{\rm i}(\dok+1)P_N(\dok+1),
\label{recursion}
\eeq
valid for $\dok=-N+1,-N+3,\ldots,N-1$ (which are the only values of $\dok$
that can occur) and with the convention that $P_N(-N)=P_N(N)=0$.
We substitute (\ref{pNupdown}) in (\ref{recursion}) and set
\beq
x=\frac{\dok}{N^{1/2}}\,, \qquad P_N(\dok)=\frac{1}{N^{1/2}}\,
\Pi_N\left( \frac{\dok}{N^{1/2}}\right),
\label{defxPi}
\eeq
expecting that in the large-$N$ limit the variables $x$ and $N$ may be
treated as continuous. On the expression thus obtained
we perform a standard expansion in negative powers of $N$. The result
is the Fokker-Planck equation
\beq
\frac{\partial\,\Pi_N(x)}{\partial N} =
  \frac{3}{2}\frac{\partial\, x\Pi_N(x) }{\partial x}
+ \frac{1}{2}\frac{\partial^2\, \Pi_N(x)}{\partial x^2}\,,
\label{FP_Pi}
\eeq
of which (\ref{PiNx}) is the stationary solution, that is, the one
solving \mbox{$\partial\,\Pi_N(x)/\partial N=0$}.

We also note that the average length of the platoons
\beq
\nu = \frac{N}{\pok} = \frac{2N}{N-\dok}
\eeq
tends to $\nu = 2$ when $N$ becomes large, as $\dok$ typically
scales as $N^{1/2}$.

%%%%%%%%%%%%%%%%%%%%%%%%%%%%%%%%%%%%%%%%%%%%%%%%%%%%%%%%%%%%%%%%%%%%%%%%%%%%%%

\end{document}